\documentclass[12pt,preprint]{aastex}

\usepackage{amsmath,amssymb} 
\usepackage{color} 

\shorttitle{Dissipation on hot Jupiters}
\shortauthors{Goodman}

\newcommand{\bez}{\boldsymbol{\hat e}_z}
\newcommand{\pd}{\partial}
\newcommand{\ud}{{\rm d}}

\newcommand{\sech}{\mathrm{sech}}

\newcommand{\vv}{\boldsymbol{v}}  

\newcommand{\vx}{\boldsymbol{x}}

\newcommand{\bdot}{\boldsymbol{\cdot}}
\newcommand{\cross}{\boldsymbol{\times}}
\newcommand{\grad}{\boldsymbol{\nabla}}

\newcommand{\diver}{\boldsymbol{\nabla\cdot}}

\newcommand{\cs}{c_{\rm s}} 
\newcommand{\BV}{Brunt-V\"ais\"al\"a\ }

\newcommand{\dy}{{\rm\,d}}

\newcommand{\opunit}{{\rm\,cm^{2}\,g^{-1}}}
\newcommand{\K}{{\rm\,K}}
\newcommand{\kms}{{\rm\,km\,s^{-1}}}

\newcommand{\kB}{k_{\rm B}}

\begin{document}
\title{Circulation and dissipation on hot Jupiters}


\author{J. Li and J. Goodman}
\affil{Princeton University Observatory, Princeton, NJ 08544}
\email{jeremy@astro.princeton.edu}

\begin{abstract}
  Many global circulation models predict supersonic zonal winds and
  large vertical shears in the atmospheres of short-period jovian
  exoplanets.  Using linear analysis and nonlinear local simulations,
  we investigate hydrodynamic dissipation mechanisms to balance the
  thermal acceleration of these winds.  The adiabatic Richardson
  criterion remains a good guide to linear stability, although thermal
  diffusion allows some modes to violate it at very long wavelengths
  and very low growth rates.  Nonlinearly, wind speeds saturate at
  Mach numbers $\approx 2$ and Richardson numbers $\lesssim 1/4$ for a
  broad range of plausible diffusivities and forcing strengths.
  Turbulence and vertical mixing, though accompanied by weak shocks,
  dominate the dissipation, which appears to be the outcome of a
  recurrent Kelvin-Helmholtz instability.  An explicit shear
  viscosity, as well as thermal diffusivity, is added to ZEUS to
  capture dissipation outside of shocks.  The wind speed is not
  monotonic nor single valued for shear viscosities larger than about
  $10^{-3}$ of the sound speed times the pressure scale height.
  Coarsening the numerical resolution can also increase the speed.
  Hence global simulations that are incapable of representing vertical
  turbulence and shocks, either because of reduced physics or because
  of limited resolution, may overestimate wind speeds.  We recommend
  that such simulations include artificial dissipation terms to
  control the Mach and Richardson numbers and to capture mechanical
  dissipation as heat.
\end{abstract}




\goodbreak
\section{Introduction}
Strongly irradiated extrasolar giant planets---``hot Jupiters''---are
expected to rotate synchronously with their orbits but to have strong
longitudinal winds that redistribute heat from the day to the night
side.  The efficiency of redistribution is important for direct
observables including infrared phase curves, the depth of secondary
eclipses in transiting systems (i.e., eclipses of the planet by its
star), and planetary spectra.  Turbulence associated with the winds
may contribute to chemical mixing of the atmosphere
\citep{Spiegel_Silverio_Burrows09}, and might even inject heat into
the convective interior of the planet \citep{Guillot_Showman02},
thereby perhaps explaining why some of these planets have lower
densities than expected by standard evolutionary models at their
estimated ages.

In a previous paper \citep[hereafter Paper I]{Goodman2009}, one of us
has argued that these thermally driven winds can be understood as
natural heat engines, which convert a fraction of the thermal power
into mechanical work: namely, the work expended to accelerate the
wind.  As with any other heat engine, the continual production of work
must be balanced by mechanical dissipation, else the kinetic energy in
the winds would grow without bound.  Paper I offered a brief
discussion of possible dissipation mechanisms.  Here we explore those
mechanisms in greater detail.  The goal is to lay the foundation for
subgrid models of the dissipative process suitable for use by global
circulation codes.

The outline of this paper is as follows.  \S\ref{sec:overview} gives
an overview of the dissipation mechanisms we
consider. \S\ref{sec:linear} presents a linear analysis of
hydrodynamic Kelvin-Helmholtz instabilities of thermally stratified
shear flows.  We show that such flows can be destabilized by thermal
diffusivity even at Richardson numbers greater than the well-known
critical value $Ri_{\rm crit}=1/4$.  Under conditions relevant to
exoplanet winds, however, we estimate that the associated growth rates
are too slow and the longitudinal wavelengths too long to be important
for dissipation, unless $Ri<1/4$, which requires transsonic flow
unless the vertical width of the shear layer is small compared to a
pressure scale height.  Hence shocks may be as important as
shear-driven turbulence for dissipation.  To investigate this, as
described in \S\ref{sec:nonlinear}, we have used the ZEUS code in two
dimensions to simulate a part of the atmosphere with horizontal and
vertical dimensions comparable to the pressure scale height.  Thermal
diffusion and viscosity have been added to the code, and thermal
driving of the wind is simulated by a horizontal body force with
nonzero curl.  We study the velocity and dissipation rate of the wind
in a statistical steady state as a function of the strength of the
driving compared to the acceleration of gravity.
\S\ref{sec:discussion} discusses our results in the context of
previous work on winds in jovian planets both within and beyond the
solar system.  \S\ref{sec:conclusions} summarizes our main
conclusions.

\section{Overview of dissipation mechanisms}\label{sec:overview}

We begin by reviewing the physical conditions in those parts of the
atmosphere where strong winds may occur.
Because of the strong stellar irradiation,
the temperature should be approximately constant with depth,
\begin{equation}
  \label{eq:temp}
  T\approx 1300 \left(\frac{f
      L_*}{L_\sun}\right)^{1/4}\left(\frac{a}{10 R_\sun}\right)^{-1/2}\K
\end{equation}
where $L_*$ is the stellar luminosity, $a$ the orbital semi-major
axis, and $f$ a dimensionless factor of order unity that depends upon
the local zenith angle of the star, the albedo, and the efficiency of
lateral heat redistribution ($f=1$ for complete redistribution and
negligible albedo).  For a solar-mass primary, the orbital period is
$P_{\rm orb}=3.66(a/10 R_\sun)^{3/2}\dy$.  The sound speed in an
atmosphere dominated by molecular hydrogen is $\cs\approx
2.75(T/1300\K)^{1/2}\kms$.  A characteristic circulation timescale is
\begin{equation}
  \label{eq:tcirc}
  t_{\rm circ}= \frac{2\pi R_P}{\cs}\approx 1.85\left(\frac{T}{1300\K}\right)^{-1/2}\left(\frac{R_p}{R_J}\right)\dy\,,
\end{equation}
where $R_p$ is the planetary radius.  Assuming that the planet rotates
synchronously, by this definition $t_{\rm circ}$ is comparable to the
rotation period, as it is on Earth.  Coriolis effects are likely to be
important but not as dominant as on a more rapidly rotating and colder
planet such as Jupiter itself.

Several considerations point to transsonic wind speeds, as discussed
in Paper I.  First, absent dissipation, flow along isobars
from the day to the night side, starting from rest, would accelerate to
\begin{equation*}
U\sim \sqrt{g\Delta z}\ln\frac{T_{\rm day}}{T_{\rm night}}\,,
\end{equation*}
where $\Delta z\sim H_p=\cs^2/\gamma g$ is the range of depths over which
the thermal forcing changes sign [see eq.~\eqref{eq:battery}].
Indeed, repeated cycling between day and
night could accelerate the wind indefinitely, were there no
mechanical dissipation.  Second, the strong entropy stratification
tends to stabilize the flow against
turbulent dissipation at small Mach numbers.  And third, many
numerical models of exoplanetary circulation do predict supersonic
winds (see Paper I for references).

Another direct consequence of the radiatively induced temperature
\eqref{eq:temp} is that the vertical density profile is approximately
exponential, with a scale height $H_p=\kB T/gm_{H_2}$ that is small compared to the
planetary radius:
\begin{equation}
  \label{eq:scaleheight}
  \frac{H_p}{R_p}\approx 2.9\times
  10^{-3}\left(\frac{T}{1300\K}\right) \left(\frac{R_p}{R_J}\right)^{-1}\left(\frac{g}{g_J}\right)^{-1}\,,
\end{equation}
where $g=GM_p/R_p^2$ is the surface gravity and $g_J\equiv
GM_J/R^2_{J,\rm eq}\approx 2500\,{\rm cm\,s^{-2}}$.

The pressure at the infrared photosphere is
\begin{equation}
  \label{eq:Pphot}
p_{\rm ph}=\frac{2g_p}{3\kappa_{\textsc{r}}}  \approx 1.7
\left(\frac{\kappa_{\textsc{r}}}{\opunit}\right)^{-1}\left(\frac{g}{g_J}\right){\rm\,mbar}\,,
\end{equation}
where $\kappa_{\textsc{r}}$ is the Rosseland mean opacity appropriate to the temperature \eqref{eq:temp}.
A larger characteristic pressure is that at which the thermal time is
comparable to $t_{\rm circ}$.  Taking $t_{\rm th}=H_p^2/\chi$, where 
$\chi= 16\sigma T^3/\kappa\rho^2 c_P$ is the thermal diffusivity and
$c_P\approx 3.5\kB/m_{H_2}$ the specific heat capacity, we estimate the
latter pressure at
\begin{equation}
  \label{eq:Pth}
  p_{\rm th}\approx 70\left(\frac{\kappa_{\textsc{r}}}{\opunit}\right)^{-1/2}\left(\frac{T}{1300\K}\right)^{5/4}
\left(\frac{g}{g_J}\right)\left(\frac{R_p}{R_J}\right)^{1/2}{\rm\,mbar}\,.
\end{equation}

The Reynolds number of these winds is enormous.  The
dynamic viscosity of molecular hydrogen is \citep{Stiel_Thodos63}
$\rho\nu=1.9\times10^{-4}(T/1000\K)^{0.65}{\rm g\,cm^{-1}\,s^{-1}}$,
and therefore
\begin{equation}
  \label{eq:Re_est}
  Re\equiv \frac{\cs H_p}{\nu}\approx 5\times
  10^8\left(\frac{p}{\,{\rm mbar}}\right)
\left(\frac{T}{1300\K}\right)^{-0.15}\left(\frac{g}{g_J}\right)^{-1}\,.
\end{equation}
At pressures above a microbar at most, therefore, it is appropriate to regard
the winds as inviscid.  

By other dimensionless measures, however, these flows are less ideal.
At these temperatures and pressures, thermal ionization of alkali
metals yields magnetic Reynolds numbers $Rm\equiv\cs H_p/\eta\sim
O(1)$, where $\eta$ is the magnetic diffusivity.
\citet{Batygin_Stevenson10} and \citet{Perna_Menou_Rauscher10} have
suggested that non-ideal MHD effects could exert a significant drag on
the winds and perhaps even contribute to heating the convective
interior in the presence of a background planetary magnetic field
$\gtrsim 10\,{\rm G}$.

A dimensionless inverse measure of thermal (rather than magnetic)
diffusion is the Peclet number
\begin{equation}
  \label{eq:Pecletdef}
  Pe\equiv\frac{\cs H_p}{\chi}\approx 0.2\left(\frac{p}{\rm
      mbar}\right)^2
\left(\frac{\kappa_{\textsc{r}}}{\opunit}\right)\left(\frac{T}{1300\K}\right)^{-7/2}
\left(\frac{g}{g_J}\right)^{-1}\,.
\end{equation}
Thus, at the pressure \eqref{eq:Pth} where thermal and advection
timescales may be comparable, $Pe\sim 10^3$, with no direct dependence
on $\kappa_{\textsc{r}}$.

Another important dimensionless quantity is
the Richardson number, which characterizes the relative strength of
stratification and shear.
The local Richardson number at altitude $z$
is $Ri(z)\equiv N^2/(dU/dz)^2$, where
$N(z)$ is the \BV frequency [eq.~\eqref{eq:BV}].
In the limit of infinite $Re$ and $Pe$, a
necessary condition for incompressible shear-driven instability is that $Ri<1/4$
somewhere in the flow \citep{Drazin_Reid81}.
For a flow that changes smoothly by $\Delta U$ over a vertical distance
$\Delta z$, a characteristic value for $Ri$ in a
constant-temperature background is, using eq.~\eqref{eq:BV} and noting
$H_p=\cs^2/\gamma g$,
\begin{equation}
  \label{eq:Riest}
  Ri \approx\frac{(\gamma-1)}{M^2\gamma^2}\left(\frac{\Delta z}{H_p}\right)^2\,,
\end{equation}
where $M= 2\Delta U/\cs$ is the Mach number associated with the
amplitude of the shear profile.  Thus since $\gamma\approx 7/5$ for an
atmosphere dominated by molecular hydrogen, $M\gtrsim 0.9 (\Delta
z/H_p)$ is required for instability.  This linear stability criterion
can be related to an energy principle and therefore should govern
nonlinear instabilities also.  However, when the stratification is
primarily thermal rather than compositional, as we expect for the wind
zone of these atmospheres, its stabilizing influence can be undercut
by diffusion of heat.  Thus instability may be possible at $Ri>1/4$ if
the Peclet number \eqref{eq:Pecletdef} is not too large.  One of the
goals of the present study is to quantify this statement for
hot-Jupiter winds.

From this survey of characteristic physical conditions and
dimensionless numbers, three possible sources of mechanical
dissipation for the winds come to the fore: (1) MHD effects, probably
requiring $B\gtrsim 10\,{\rm G}$; (2) shear-driven turbulence,
requiring $Ri<1/4$ and/or moderate $Pe$; and (3) shocks, requiring $M>
1$ as we have defined it.  Another possibility, raised in Paper~I, is
the double-diffusive Goldreich-Schubert-Fricke instability
\citep{GoldreichSchubert67,Fricke68}, which we do not consider here
for two reasons: first, although the linear GSF instability is
axisymmetric, we believe that its saturation can be studied reliably
only with high-resolution 3D simulations, whereas we limit ourselves
to 2D; and second, because the instability shuts off when (in the
usual cylindrical coordinates) $\partial v_{\phi}/\partial z=0$ and
$\partial(\varpi v_\phi)^2/\partial\varpi\ge0$, which conditions are
not inconsistent with strong vertical shears at equatorial latitudes.
Nonlinear MHD effects present even greater numerical challenges.
Therefore, we concentrate on Richardson turbulence and shocks.

\section{Linear Analysis}\label{sec:linear}
We are interested in the hydrodynamic stability of a horizontal shear
flow.  We first consider the Boussinesq limit in which we neglect
density variations except where coupled to gravity.  The equations of
hydrodynamics become
\begin{subequations}\label{eq:basic}
  \begin{align}
  \label{eq:NavierStokes}
\frac{\ud\vv}{\ud t}&= -\grad\tilde p +\theta\bez+\nu\nabla^2\vv,\\
\label{eq:thetaeqn}
\frac{\ud\theta}{\ud t}&= -N^2\bez\bdot\vv +\chi\nabla^2\theta,\\
\label{eq:cont}
\diver\vv &=0.
\end{align}
\end{subequations}
Here kinematic viscosity $\nu$ is taken to be a constant,
\begin{equation}
  \label{eq:thetadef}
  \theta\equiv -g\left(\frac{\pd\ln\rho}{\pd\ln T}\right)_p\frac{\delta T}{\bar T}
\to g\frac{\delta T}{\bar T}\qquad\mbox{(ideal gas)}
\end{equation}
is proportional to the fractional temperature variation,
\begin{equation}
  \label{eq:BV}
  N^2\equiv\frac{\rho g^2}{p}\left[\left(\frac{\pd\ln T}{\pd\ln P}\right)_S
-\frac{\ud\ln T/\ud z}{\ud\ln P/\ud z}\right]_0
\to\frac{\gamma-1}{\gamma}\frac{g}{H_p}\qquad\mbox{(ideal gas \& $T_0=$constant)}
\end{equation}
is the square of the \BV frequency of the background hydrostatic
state, and the pressure scale height
\begin{equation}
  \label{eq:Hp}
  H_p\equiv \left|\frac{\ud\ln p}{\ud z}\right|_0^{-1}\to\frac{\kB
    T_0}{\mu m_p g}=\frac{\cs^2}{\gamma g}\,.
\end{equation}
We assume the optically thick diffusion limit for radiative transfer
in which $\chi=16\sigma T^3/3 \kappa c_p \rho^2$.  We further pick the
opacity $\kappa \propto T^3\rho^{-2}$ so that $\chi=\text{constant}$.
Note that in our energy equation we assume no background temperature
gradient.  Other authors have studied the linear stability of the
Boussinesq equations \citep[see
e.g.,][]{Gage1973,Dudis1974,Maslowe1971,Koppel1964,Miller_Gage1972,Gage1972},
but they do not consider in detail the destabilizing effects of
thermal diffusivity in the viscous, optically thick limit.  We are
interested in the growth rates in the weak viscosity regime and their
relevance to our numerical simulations.

Assuming a plane-parallel background state with horizontal flow
\begin{equation}
  \label{eq:zeroth}
  \theta_0=0=\tilde p_0,\qquad\vv_0=U(z)\boldsymbol{\hat e}_x,
\end{equation}
we can expand all perturbations with the dependence $\exp(-i\omega
t+ik_x x+ik_y y)$ to obtain the linearized $6^{th}$ order system of
equations
\begin{subequations}\label{eq:6system}
\begin{align}
\label{eq:NavierStokeslin}
v_{1z}''''&=\left(\frac{ik^2\sigma}{\nu}-k^4
-\frac{ik_xU''}{\nu}\right)v_{1z}
+\left(2k^2-\frac{i\sigma}{\nu}\right)v_{1z}''+\frac{(k^2)}{\nu}\theta_1,\\[1ex]
\label{eq:thetaeqnlin}
\theta_1''&=\left(-\frac{i \sigma}{\chi}+k^2\right)\theta_1+\frac{N^2}{\chi}v_{1z},
\end{align}
 \end{subequations}
where $\sigma=\omega-k_xU$, $k\equiv(k_x^2+k_y^2)^{1/2}$,
and primes denote $d/dz$.

We use a relaxation method to solve this system of equations, and we
test our results against limiting cases of background velocity profile
$U$ and parameters $\nu,\chi,$ and $N^2$ for which there are simple
analytic solutions.  See the Appendix for the limits and analytical
solutions.  We can nondimensionalize our parameters by defining the
Richardon number as $Ri=N^2H_p^2/U_{max}^2$,
Reynolds number $Re=H_pU/\nu$, Peclet number $Pe=H_pU/\chi$, and
Prandtl number $Pr=\nu/\chi=Pe/Re$.  (In \S2, we scaled $Re$ and $Pe$
by the sound speed rather than the flow velocity in the expectation of
a Mach number of order unity, but that is not appropriate in the
present Boussinesq context.)  We give all our linear results in terms
of these dimensionless numbers.

For the problem of interest we specify boundary conditions for
the system \eqref{eq:6system} as follows
We impose $v_{1z}=0$ at $z=\pm z_{max}$ so
that the fluid perturbations do not penetrate the boundary.  We
further impose $v_{1z}''=-ik_xv_{1x}'-ik_yv_{1y}'=0$,
corresponding to vanishing viscous stress, and $\theta_1=0$,
corresponding to fixed temperature, at both boundaries.

Consider now the limit $\nu \to 0,
\chi \to 0$ of eqns~\eqref{eq:6system}
\begin{equation}
v_{1z}''=\left(-\frac{k_xU''}{\sigma}+k^2-\frac{\left(k^2\right)N^2}{\sigma^2}\right)v_{1z}.
\end{equation}
From WKB arguments, the time taken for a wavelike disturbance to reach
$z=0$ from $z=z_1\neq0$ is infinite if $Ri > 1/4$.  Further, from
energy arguments the potential energy barrier separating two fluid
elements at different heights exceeds the available kinetic energy
that can be liberated if $Ri>1/4$ \citep{Chandrasekhar1961}.  From these
considerations, we expect the stratified shear flow to be stable for
$Ri \gtrsim 1/4$ in the inviscid case with no thermal diffusion.  As
we turn up the thermal diffusivity the potential energy barrier
between fluid elements at different heights can be more easily
overcome, and we expect instabilities to survive to higher Richardson
number.  We expect there to be a critical Richardson number, above
which the stratified flow is stable and below which there exist
unstable Kelvin-Helmholtz like modes.  We refer to this critical
solution as the marginally stable mode.

\begin{figure}[htp]
\centering
\includegraphics[scale=.4]{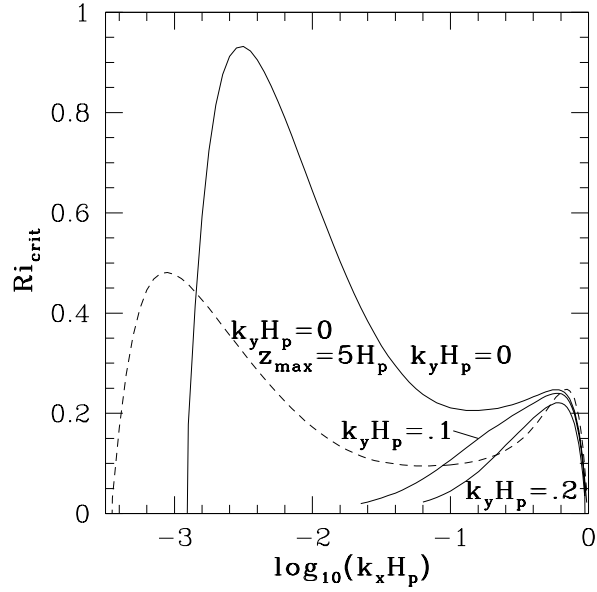}
\caption{Marginally stable Richardson number as a function of $k_xH_p$
for $Pe=10,Re=10^{3}$, and at a sequence of $|k_yH_p|=0,.1,.2$.  As
$|k_yH_p|$ increases the critical Richardson number decreases at all
$k_xH_p$.  The dashed curve shows a separate run at $z_{max}=5H_p$,
indicating the long wavelength results have not
converged.}\label{fig:kxRiky}
\end{figure}

For our stability analysis we pick background velocity profile
$U=U_0\tanh(z/H_p)$.  Squire's theorem tells us that in the limit of
vanishing thermal diffusivity the largest growth rates are obtained
for $k_yH_p=0$ \citep{Drazin_Reid81}, so we might expect this result to
generalize to the case with thermal diffusion.  In Fig
\ref{fig:kxRiky} we show the marginally stable Richardson number as a
function of $k_xH_p$ for $k_yH_p=0,.1,.2$ and $Pe=10,Re=10^{3}$.  As
$|k_yH_p|$ increases, the critical Richardson number decreases for all
$k_xH_p$.  Growth rates at fixed Richardson number correspondingly
decrease.  We set $k_yH_p=0$ hereafter because we are interested in
the fastest growing modes.  Note that we find two distinct
instabilities at long and short wavelengths.  The long wavelength
instabilities have {\it not} converged with $z_{max}$.  The solid lines
indicate solutions computed with $z_{max}=3H_p$, and the dashed line
indicates a solution computed at $z_{max}=5H_p$.  The long wavelength
instability is driven to longer wavelengths as $z_{max}$ increases.
We run into similar issues encountered by \cite{Dudis1973} in the long
wavelength limit.  We must integrate to larger and larger $z_{max}$ or
use an appropriate asymptotic boundary condition as $k_xH_p \to 0$ to
obtain better convergence.

Although our code has not converged for long wavelengths, our results
suggest that there may be a real inviscid instability in the limit
$k_xH_p \to 0$.  This is demonstrated analytically in Appendix B.  The
dimensionless form of the equations of motion given in the Appendix
suggest that the long-wavelength modes should be sensitive to the
combination of parameters $Pe Ri\,k\,z_{\max}^2/H_p$, and the analysis
guarantees the existence of the long-wavelength modes only where this
combination is $\ll 1$; some version of the long-wavelength modes may
exist more generally, but we have not proven it.  So it is not
surprising that our numerical results for these modes have not
converged with respect to $z_{\max}$.

\begin{figure}[htp]
\centering
\includegraphics[scale=.4]{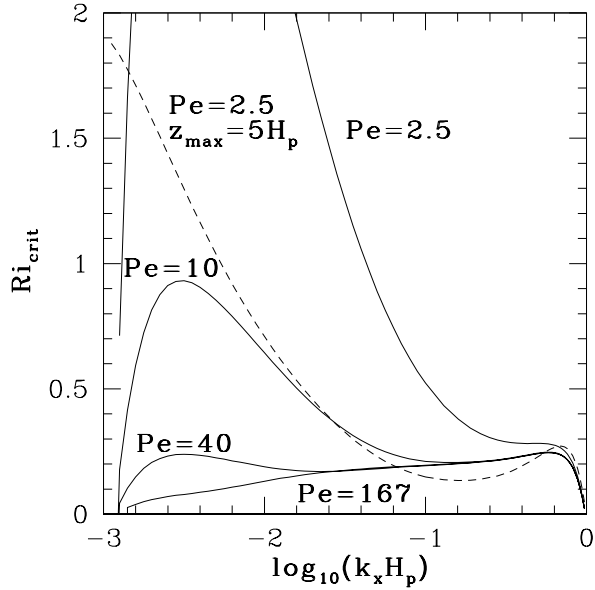}
\caption{Marginally stable Richardson number vs. wavenumber
$k_xH_p$, for fixed Reynolds number $Re=10^{3}$ and a
sequence of Peclet numbers $Pe\equiv(\nu/\chi)Re$.  At $Pe\lesssim40$,
a long-wavelength instability that survives at $Ri > 1/4$ develops.
At $Pe\lesssim2.5$, the short-wavelength
modes destabilize above $Ri=1/4$.  The dashed curve indicates a run at
higher $z_{max}=5H_p$.  The short-wavelength results have converged
reasonably well, and for $Pe\sim 10^2-10^4$, the critical Richardson
number remains fixed at $Ri=1/4$.}\label{fig:kxRi}
\end{figure}

In Fig \ref{fig:kxRi} we show the critical Richardson number as a
function of $k_xH_p$ at $Re=10^{3}$ and at a sequence of Peclet
numbers.  As the Peclet number decreases below $\sim 40$, the long
wavelength instability begins to strengthen and survive above $Ri=1/4$.
The dashed line is a run at higher $z_{max}=5H_p$ and shows the
convergence of short wavelength results.  As the Peclet number
decreases below $\sim 2.5$ the flow at short wavelengths begins to
destabilize above $Ri=1/4$.  For high Reynolds number and high Peclet
number our short wavelength results are in rough agreement with the
inviscid results of Dudis (1974).  There are quantitative differences
due to our assumption of an isothermal background temperature profile.

\begin{figure}[htp]
\centering
\includegraphics[scale=.4]{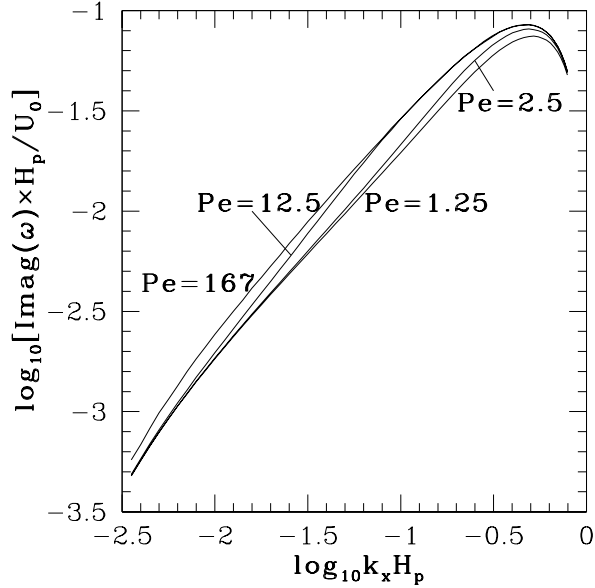}
\caption{Growth rates of modes with $Ri=\frac12Ri_{\rm crit}$ as a
function of $k_xH_p$, for fixed $Re=10^{3}$ and
$Pe=1.25,2.5,12.5,167$.  The long wavelength instability that survives
to high Richardson number has growth rates over two orders of magnitude
weaker than shorter wavelength growth rates.}\label{fig:kxOm}
\end{figure}

In Fig \ref{fig:kxOm} we show the growth rates of the modes with
$Ri=\frac12Ri_{\rm crit}$ as a function of $k_xH_p$, for fixed $Re=10^{3}$
and at a sequence of Peclet numbers.  The long wavelength growth rates
at $k_xH_p=3.2/Re$ are well over two orders of magnitude smaller than
shorter wavelength growth rates.  This result agrees with our
estimation that the inviscid instability in the limit $k_xH_p \to 0$
has weak growth rate.  For our numerical simulations we will be
primarily interested in the more unstable short wavelength
instabilities that have maximum growth at $k_xH_p\sim0.5$.  Note that
$Ri_{\rm crit}$ increases with decreasing $Pe$, so we are computing growth
rates at higher Richardson number for the lower Peclet number curves.
At fixed $Ri$, growth rates increase with decreasing $Pe$, as we
expect from simple heat diffusion arguments.  

We conclude from Figs \ref{fig:kxRi} and \ref{fig:kxOm} that the
linear growth rates of the fastest growing mode are unchanged in the
parameter regime we are interested in, $Pe \sim 10^2-10^4$.

\newpage
\section{Numerical Simulations}\label{sec:nonlinear}
We use ZEUS-2D v2.0 \citep{Stone1992} to run hydrodynamical
simulations of a horizontal wind profile in a local equatorial section
of the atmosphere.  Cartesian coordinates $x$ and $z$ correspond to
longitude and altitude, respectively, each extending over a few
pressure scale heights: typically $20 H_p\times 5H_p$.  Longitude
$x$ is periodic, though with a period much less than the
true circumference of the planet.  The simulations include vertical
gravity, thermal diffusion of heat, shear viscosity, and a horizontal
body force per unit volume $\rho
f(z)\boldsymbol{\hat e}_x$ [eqs. \eqref{eq:NavierStokesNumerical} \&
\eqref{eq:forcing}] rather than thermal forcing.  A few words of
justification are in order.

The pressure acceleration is
nonpotential due to longitudinal entropy gradients; the component of its curl
normal to the equatorial plane is, in spherical coordinates $r\theta\phi$,
\begin{equation}\label{eq:battery}
\boldsymbol{\hat e}_\theta\cdot\left(\grad \rho^{-1}\cross\grad p\right)
= \frac{1}{r\rho^2}\left(\frac{\partial\rho}{\partial T}\right)_p
\left[\frac{\partial T}{\partial\phi}\frac{\partial p}{\partial r}-
        \frac{\partial p}{\partial\phi}\frac{\partial T}{\partial r}\right]
\approx \frac{g}{r}\left(\frac{\partial\ln T}{\partial\phi}\right)_p\,,
\end{equation}
in which the final form uses the ideal-gas law and assumes approximate vertical hydrostatic
equilibrium.
This curl  must vanish in the (nearly isentropic) convection zone.
Furthermore, since thermal forcing by itself does not add any net momentum to the 
flow,\footnote{\citet{Arras_Socrates10} have suggested, however, that in conjunction with
the tidal field of the host star, thermal forcing may lead to net torques on the planet.}
the nonpotential longitudinal force must reverse sign with depth within the radiative atmosphere.
In fact one expects this nonpotential ``battery'' to be concentrated at
pressures $\lesssim p_{\rm th}$ [\eqref{eq:Pth}], which
is quite shallow compared to the radiative-convective interface, typically $\sim$kbar.
The nonpotential force must also reverse sign at least twice in longitude because \eqref{eq:battery}
is periodic in $\phi$, but the horizontal scale of variation of this force is probably small
compared to its vertical scale since the radiative atmosphere is thin.
Hence the profile \eqref{eq:forcing}, which is horizontally constant and extends
over $\sim 2 H_p$ vertically, reversing sign within this depth, is a plausible representation of the
horizontal forcing within our local computational domain.

\subsection{Setup}
The source terms
for momentum and energy in ZEUS are modified by the addition of the
terms within square brackets below:
\begin{equation}
  \label{eq:NavierStokesNumerical}
\rho\left(\frac{\partial\vv}{\partial t}\right)_{\rm sources}= -\grad p -\rho g \bez -\nabla \cdot \boldsymbol{Q}+
\left[\rho f(z)\boldsymbol{\hat e}_x + \boldsymbol{\nabla\cdot\sigma'}\right]
\end{equation}
and
\begin{equation}
  \label{eq:energy}
\left(\frac{\partial e}{\partial t}\right)_{\rm sources}= -p
\boldsymbol{\nabla\cdot v} -\boldsymbol{Q\cdot\nabla v} +
\left[\grad\cdot (\chi \rho c_p \grad T) + \boldsymbol{\sigma'\cdot\nabla v}\right]\,,
\end{equation}
in which $\boldsymbol{\sigma'}$ is the viscous shear tensor, with components
\begin{equation*}
\sigma'_{ik}=\nu \rho\left(\frac{\pd v_i}{\pd x_k}+\frac{\pd
v_k}{\pd x_i}-\frac{2}{3}\delta_{ik}\frac{\pd v_l}{\pd x_l}\right)\,,
\end{equation*}
where $\nu\text{ and } g$ are taken to be constants.  ZEUS' standard
von Neumann \& Richtmyer artificial viscosity tensor $\boldsymbol{Q}$
is applied with shocks spread over $\approx 2$ zones.  The thermal
diffusivity $\chi$ is again given in the optically thick limit by
$\chi=16\sigma T^3/3 \kappa c_p \rho^2={\rm const}$; constant $\chi$
would correspond therefore to $\kappa\propto T^3\rho^{-2}$.  The
horizontal forcing profile is
\begin{equation}
  \label{eq:forcing}
f(z)=\epsilon g\left[2\sech^2(z/H_p)\tanh(z/H_p)+\alpha \sech^2(z/H_p)\right],
\end{equation}
where $H_p$ is the pressure scale height, $\alpha$ is a parameter that
we adjust to ensure zero net momentum input on our grid, and
$\epsilon\ll 1$ determines the overal strength of the forcing.  This form
is motivated by the viscous acceleration on a fluid with hydrostatic
density profile and horizontal velocity profile given by
$v_x=U_0\tanh(z/H_p)$.  The viscous acceleration in this case is given
by $\epsilon g=\nu U_0/H_p^2$ and $\alpha=1$.  We use an FTCS
differencing scheme for our modified source terms, and the ZEUS
timestep constraint must satisfy $\delta t<\delta t_\nu=(\Delta
x)^2/4\nu$ and $\delta t<\delta t_\chi=(\Delta x)^2/4\chi$ for
stability.  We are interested in $Pr<1$, so the latter constraint is a
tighter restriction on $\delta t$.  But $\delta t_\chi/\delta t_{CFL}
\approx Pe/4N_p$, where $N_p$ is the number of grid points per pressure
scale height and $Pe$ is the Peclet number in the flow.  Dimensionless
parameters for our numerical results are scaled to the sound speed
$c_s$ as in \S\ref{sec:overview}.  Our simulations generally satisfy
$Pe\sim 10-100$ and $N_p\sim O(10)$, so $\delta t_\chi/\delta
t_{CFL} \sim O(1)$ and the diffusion and CFL timesteps are usually
comparable.

As stated, $x$ is periodic.  At $z=z_{\min}$ and $z_{\max}$,
we impose reflecting conditions for density and velocity
and constant temperature $T=T_{\rm wall}$.  We initialize the
density to $\rho(x,z)=\rho_0$ and the internal energy density to
$e(x,z)=e_0$, allowing the flow to settle under the influence of
gravity.  The velocity is initialized to
$\vec{v}=U_0\tanh(z/H_p)\boldsymbol{\hat e}_x$.  The wall temperature
is set to the initial grid temperature, $T_{\rm wall}=T_0=(e_0/\rho_0)\mu
m_p(\gamma-1)/k_B$.  We pick $\rho_0=1, e_0=.01, \gamma=1.4,$ and
$g=.004$.  Our simulations are run over 5 vertical pressure scale
heights, from $z=-2.5H_p \text{ to } z=2.5H_p$.  The density at the
base of our simulation if it were in hydrostatic equilibrium is then
$\rho_{base}=5\rho_0$.  The code units for our problem are given in
table \ref{tab:units}.  We nondimensionalize parameters according to
these units.

\begin{deluxetable}{ccccc}
\tablewidth{0pt}
\tablecaption{Code Units.}
\tablehead{\colhead{Unit} & \colhead{Length} & \colhead{Time} & \colhead{Density} & \colhead{Acceleration}} 
\startdata
Value & $H_p=1$ & $t=H_p/c_s=13.36$ & $\rho_{base}=5$ & $g=.004$ \\ \hline
\enddata
\label{tab:units}
\end{deluxetable}

We run our simulations over 20 scale heights in the horizontal
direction.  We are physically interested in $Pe^{-1}\approx
10^{-2}-10^{-4}$, $Re^{-1} \to 0$, but the viscosity must be large enough
to maintain energy conservation, and $\epsilon \le 10^{-2}$.  In the
absence of a shear viscosity the grid loses energy to numerical
dissipation at the grid scale.  We must pick $\nu$ large enough such
that the viscous dissipation length $l_{\nu}\sim(|\pd_j v_i|/\nu)^{-1/2}$
is larger than the grid scale, and the dissipation of solenoidal turbulence is
converted to internal energy.  Our standard runs
have $N_p=10$ grid points per pressure scale height, but for certain
runs which conserve energy less well we use up to $N_p=30$ grid points
per pressure scale height.

\subsection{Tests}\label{sec:tests}

We run two simple test problems to verify proper implementation of the
horizontal forcing and viscosity.  Using a simplified shear viscosity
$\pd v/\pd t=\nu \nabla^2 v$ and no viscous heating, there is
an exact solution for the viscous decay of initial velocity profile
given by $U(z,0)=\mbox{sign}(z){\boldsymbol{\hat e}_x}$.  The velocity is
given at later times by $U(z,t)=\mbox{erf}(z/2\sqrt{\nu
t}){\boldsymbol{\hat e}_x}$.  Again using this simplified shear
viscosity and horizontal forcing given by $f(z)=\epsilon
2\sech^2(z/H_p)\tanh(z/H_p)$, there is an exact laminar solution for the
velocity profile.  The laminar solution satisfies $0=f(z)+\nu U''(z)$,
where $U(z)=(\epsilon H_p^2/\nu)\tanh(z/H_p)$.  We conclude from
these two tests that there is negligible numerical dissipation in the
laminar regime.  We can also use the latter to estimate the wind speed
in the laminar regime when the viscous heating is implemented.

We are interested in the behavior of the solutions to eqns
\ref{eq:NavierStokesNumerical} and \ref{eq:energy} at different forcing
amplitudes.  We check that the parameter $\alpha$ has a second order
effect on the forcing and that we can roughly characterize the
amplitude of the forcing by the single parameter $\epsilon$.

\subsection{Diagnostics of dissipation}\label{sec:diagnostics}

We must also maintain energy conservation in our simulations.  We are
continually injecting energy onto the grid via the horizontal forcing,
and to reach a statistical steady state this energy must first be
converted into internal energy and then allowed to diffuse through the
boundary walls.  Because ZEUS is not based on a
conservative form of the energy equation, numerical diffusion dissipates kinetic energy
on the grid scale without converting it to internal energy;
to avoid this, it is necessary to add an explicit shear viscosity (in
addition to the artificial bulk viscosity used to mediate shocks).

We compute a fractional energy error
\begin{equation}
\label{eq:energyerror}
\frac{\dot{E}}{\dot{E}_{\rm in}}=\frac{\dot{E}_{\rm in}-\dot{E}_{\rm kin}-\dot{E}_{\rm int}-\dot{E}_{\rm grav}
-\dot{E}_{\rm out}}{\dot{E}_{\rm in}},
\end{equation}
where
\begin{align*}
\dot{E}_{\rm in}&=\iint \rho v_x f(z) dxdz  &
\dot{E}_{\rm kin}&=\frac{d}{dt}\iint\tfrac{1}{2}\rho v^2 dxdz &
\dot{E}_{\rm int} &= \frac{d}{dt}\iint e \,dxdz 
\\[1ex]
\dot{E}_{\rm out} &=\int\left[-\rho c_p\chi\frac{\partial T}{\partial z}\right]_{z_{\min}}^{z_{\max}}\,dx
&\dot{E}_{\rm grav}&= \frac{d}{dt}\iint \rho g z\, dxdz &
\end{align*}

We briefly mention here a number of strategies that we use to improve
energy conservation.  The fractional energy error will tend to
decrease if we keep the amplitude of the forcing function lower so
that there is less energy injection onto the grid.  In a similar vein,
if we localize the forcing function to the shear layer as opposed to
distributing it over the entire vertical extent of the grid then there
will be less energy injection and better energy conservation.  Other
strategies include increasing the kinematic viscosity or making the
grid resolution finer in order to capture kinetic energy dissipation
above the grid scale via the explicit viscosity.  Finally, we can
increase the thermal diffusivity so as to allow excess internal energy
to diffuse off the grid.  These considerations must be balanced
against keeping the computation time reasonably low and keeping the
solutions physically relevant to the problems of interest.

We consider that a statistical steady state is reached when
$\langle \dot{E}_{\rm int}\rangle_t$, $\langle\dot{E}_{\rm
kin}\rangle_t$, and $\langle\dot{E}_{\rm grav}\rangle_t$ are all $\ll\dot E_{\rm in}$,
and when the Mach number [eq.~\eqref{eq:machnumber}] and the rms vertical velocity
$\langle v_z^2\rangle_{x,y}^{1/2}$ are constant for at least a horizontal sound crossing time,
$20H_p/c_s$.  In steady state,
$\dot{E}/\dot{E}_{\rm in} \approx (\dot{E}_{\rm in}-\dot{E}_{\rm
out})/\dot{E}_{\rm in}$.  We maintain the fractional energy error to
within ten percent.

Dissipation of kinetic energy involves production of entropy.  If $s$
is entropy per unit mass, then
\begin{align*}
\rho \frac{Ds}{Dt}&= \frac{-\boldsymbol{Q:\nabla v} +\boldsymbol{\sigma':\nabla v}}{T}
+ \chi c_p\rho\left|\grad\ln T\right|^2\ +\diver\left(\chi c_p\rho\grad\ln T\right).
\end{align*}
When integrated over our computational domain, the final term is $-\dot E_{\rm out}/T_{\rm wall}$,
$\dot E_{\rm out}$ being the rate of escape of heat.  So
\begin{equation}
\label{eq:dissipation}
\dot{E}_{\rm out}=T_{\rm wall}\int \left[T^{-1}(-\boldsymbol{Q:\nabla v} +
\boldsymbol{\sigma':\nabla v}) + \chi c_p
\rho |\grad \text{ln} T|^2\right]d^2\vx\quad+\frac{d}{dt}\iint\rho s dxdy.
\end{equation}
In statistical steady state, the time average of the last term
vanishes.

While conversion of mechanical energy to heat ultimately depends upon the
``microscopic'' processes in eq.~\eqref{eq:dissipation}, 
 we gain insight into the macroscopic processes that remove kinetic energy from 
the large-scale flow---shocks and turbulence---by
monitoring the energy change in each operator split source step.
The price paid for this approach is having to deal with a mixture of
reversible and irreversible effects.
Under adiabatic conditions, for example, the compressive heating term 
$\dot{E}_{pdV}=\int-p\diver\vv\, d^2\vx$
would be completely reversible, but shocks and thermal diffusion
make $\langle\dot{E}_{pdV}\rangle\ne0$ even in steady state.
There is also a change in gravitational
potential energy $\dot{E}_{\rm grav}=\int v_z \rho g d^2\vx$
associated with the vertical gravity source step.  In many of our runs
we find that $\dot{E}_{pdV}$ and $\dot{E}_{\rm grav}$ have large
oscillations that nearly cancel.
Using integration by parts, one can show that
\begin{equation}\label{eq:Edotsum}
  \dot{E}_{pdV}+\dot{E}_{\rm grav}= 
\iint \vv\cdot\left[\rho\grad(gz)+\grad p\right]\,dxdz\,.
\end{equation}
Thus the near cancellation suggests that our flows are close to vertical hydrostatic
equilibrium despite large vertical motions.  We find it easiest to monitor the combination
$\dot{E}_{pdV}+\dot{E}_{\rm grav}$.  The time
average of this term is largely the work done to mix the fluid vertically;
absent thermal diffusion, this could lead to an almost adiabatic
temperature-pressure profile, $\partial\ln T/\partial z\approx
-(\gamma-1)g/\cs^2$, but in steady state diffusion tends to restore the stratification.
Thus vertical mixing acts as a refrigerator---a heat engine in
reverse---absorbing mechanical work to drive heat up the temperature gradient.

Shock dissipation is dominated by the artificial bulk viscosity,
$\dot{E}_{\rm abv}=\int -\boldsymbol{Q\cdot\nabla v} d^2\vx$ but receives small
contributions also from the shear stress and from the $\dot{E}_{pdV}$ term.

Since we find that shocks are weak, we can estimate the turbulent energy
dissipation rate as
\begin{equation}
\label{eq:Edotturb}
\langle \dot{E}_{\rm turb}\rangle_t\approx\langle \dot{E}_{p\diver
v}+\dot{E}_{\rm grav}\rangle_t+\langle \dot{E}_\nu\rangle_t-\langle
\dot{E}_{\rm lam}\rangle_t,
\end{equation}
where $\dot{E}_{\nu}=\int \boldsymbol{\sigma' \cdot \nabla v}$ and
$\dot{E}_{\rm lam}=\int \nu \rho (\pd \bar{v}_x/\pd z)^2d^2\vx$.  Assuming
an effective turbulent viscosity acting on the time-averaged mean
shear profile, we estimate the turbulent dissipation rate as $\langle
\dot{E}_{\rm turb}\rangle_t=\langle \int \nu_t \rho (\pd \bar{v}_x/\pd
z)^2d^2\vx\rangle_t$.  The effective shear viscosity due to turbulence in a
flow that does not resolve the turbulent energy cascade is then
$\nu_{\rm t}=\nu\langle\dot{E}_{\rm
turb}\rangle_t/\langle\dot{E}_{\rm lam}\rangle_t$.

\subsection{Results}\label{subsec:results}
\begin{figure}[htp]
\centering
\includegraphics[scale=.5]{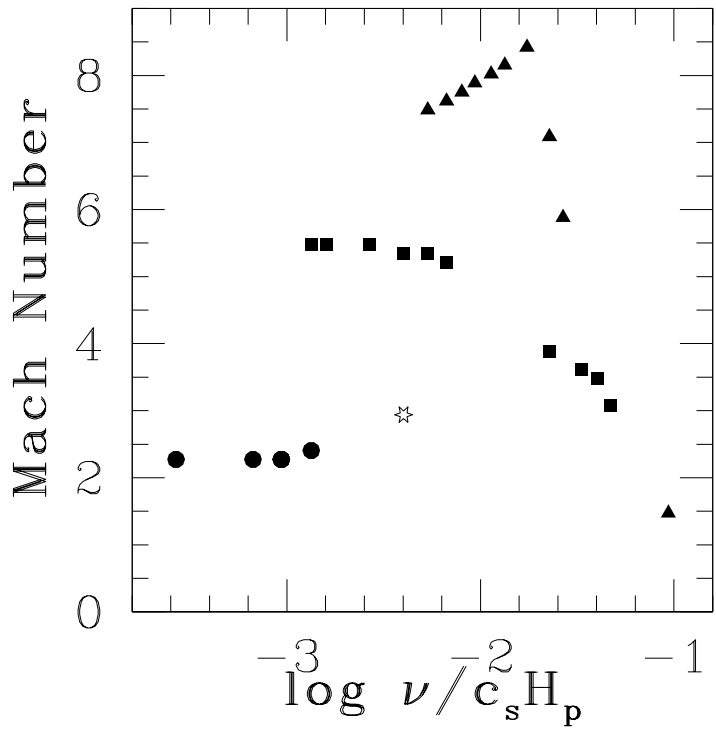}
\caption{Wind amplitude as a function of scaled viscosity $\nu/c_sH_p$ for
$\epsilon=0.1,\chi/c_sH_p=0.094$.  We find four distinct solutions in
the range $\log_{10}\nu/c_sH_p=-3 \text{ to } -1.5$, but a unique
solution, with Mach number $\mathcal{M}\sim 2.3$, in the high-$Re$
limit.}\label{eps.1g_vx}
\end{figure}

\begin{figure}[htp]
\centering
\includegraphics[scale=.5]{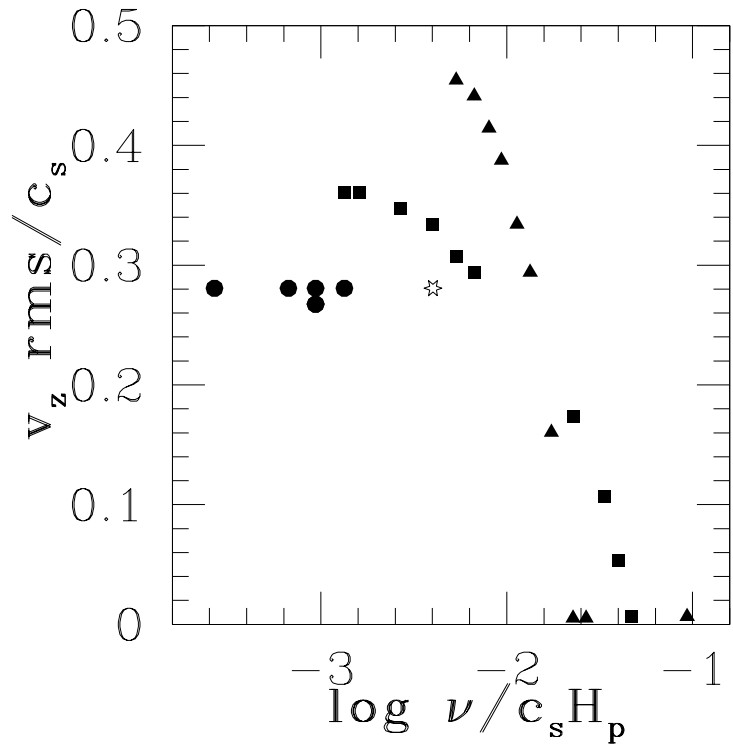}
\caption{Scaled vertical velocity fluctuations as a function of $\nu/c_sH_p$ for
$\epsilon=0.1,\chi/c_sH_p=0.094$.}\label{eps.1g_vzrms}
\end{figure}

\begin{figure}[htp]
\centering
\includegraphics[scale=.5]{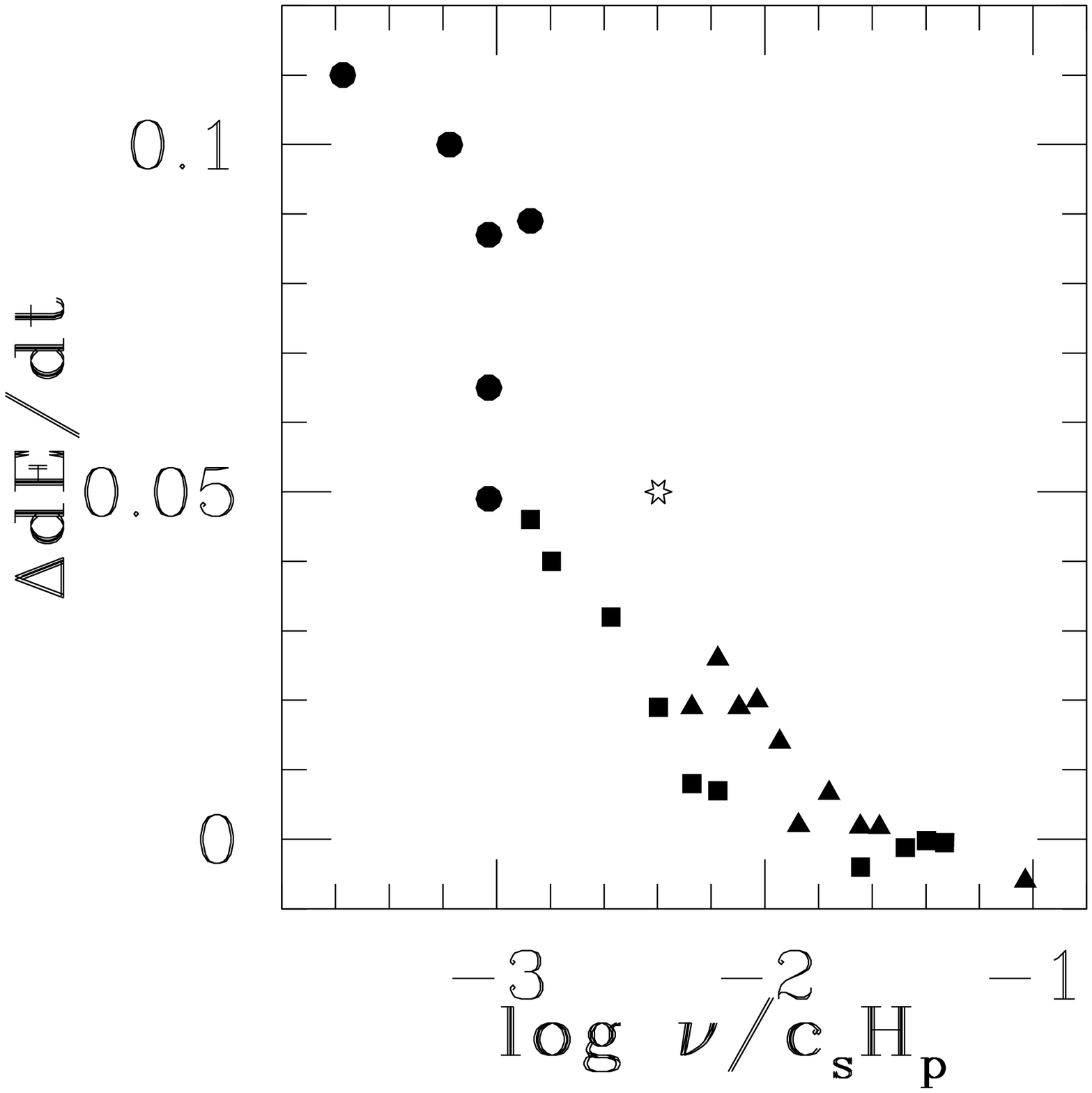}
\caption{Fractional energy error as a function of $\nu/c_sH_p$ for
$\epsilon=0.1,\chi/c_sH_p=0.094$.  The extra two lower filled circles at
$\log_{10}\nu/c_sH_p=-3.03$ demonstrate the convergence of our slow A
solutions.  The energy error decreases with increased resolution, but
the gross properties of the flow remain
unchanged.}\label{eps.1g_energy}
\end{figure}

Figs \ref{eps.1g_vx},\ref{eps.1g_vzrms}, and \ref{eps.1g_energy}
respectively give the run of Mach number, defined in
eq.~\eqref{eq:machnumber}, rms vertical velocity, and fractional
energy error versus $\nu/c_sH_p$ for $\chi/c_sH_p=0.094,\epsilon=0.1$.
The Prandtl number is less than one for all of these runs.  Note the
presence of four distinct solutions at $\nu/c_sH_p >
10^{-3}$.  There is a fast solution, denoted by filled triangles, a
medium velocity solution, denoted by filled squares, and two slow
solutions, denoted by filled circles and open stars.  Where
solution branches overlap, we trace each branch by
starting from a representative member and then gradually varying
parameters.  We use a similar parameter deformation to verify that
each branch is limited to a certain range of $\nu/c_sH_p$ as indicated
in Figs \ref{eps.1g_vx}-\ref{eps.1g_energy}.

The solutions differ in the importance of laminar,
turbulent, shock, and irreversible compressional dissipation. 
The fast solutions (triangles) have high laminar energy dissipation.
Fig \ref{eps.1g_vzrms} indicates that there is a significant vertical
component to the velocity field, but the bulk of the viscous
dissipation is laminar rather than turbulent.  This is reflected by
wavelike oscillations in the flow as opposed to turbulent
Kelvin-Helmholtz rolls.  The medium and slow B solutions have stronger
irreversible compressional heating and shocks, but they still have
relatively high laminar dissipation.  For the planetary wind problem
of interest to us this laminar dissipation is unphysical, and as we
increase the Reynolds number these solutions disappear.  We conclude
that these unphysical solutions are supported by the shear viscosity.
Numerical diffusion can also support these unphysical solutions.
At a grid resolution of $N_p=3$ grid points per pressure scale height
and parameters from Fig \ref{eps.1g_vx}, we find unphysical solutions
even at $\nu/c_sH_p=0$.  We estimate that it is necessary to have $N_p
\gtrsim O(10)$ in order to enter the
high-$Re$ regime where there exists a single, physical
solution.  This unique solution is the slow A solution in Figs
\ref{eps.1g_vx} - \ref{eps.1g_energy}.  It is characterized by weak
laminar dissipation and strong irreversible compressional and shock
dissipation.  We check the convergence of this solution with runs at
1.5 and 2 times the standard resolution.  Extra filled circles at
$\log_{10} \nu/c_sH_p=-3.03$ in figs \ref{eps.1g_vx} -
\ref{eps.1g_energy} indicate higher resolution runs.  The wind speeds
and rms vertical velocities overlap, but the energy errors progress
downwards at higher resolution.

We summarize these results in Table \ref{tab:eps.1g} by providing
quantitative diagnostics for each solution type.  We give the shear
viscosity, turbulent viscosity, relative importance of artificial bulk
viscosity, relative importance of turbulent dissipation, Mach number,
and Richardson number for a representative solution of each type.  The
laminar and turbulent energy dissipation rates are directly
proportional to the quoted shear and turbulent viscosities, as
described in \S\ref{sec:diagnostics}.  At this stage we are only
providing rough diagnostics to characterize each solution type, but we
infer that shocks contribute weakly to the irreversible compression
because the artificial-bulk-viscosity dissipations are relatively
small.  The Mach number is computed as
\begin{equation}
\label{eq:machnumber}
\mathcal{M}\equiv\max_{z}\left[\frac{\int\rho(x,z)v_x(z)dx}{c_{\rm s,wall}\int\rho(x,z)\,dx}\right]
-\min_{z}\left[\frac{\int\rho(x,z)v_x(z)dx}{c_{\rm s,wall}\int\rho(x,z)\,dx}\right]\,.
\end{equation}
We compute the minimum Richardson number in our flows via
\begin{equation*}
Ri=\min_z \frac{N^2}{(\pd \bar{U}/\pd z)^2} = \min_z\left[
\frac{\bar{p}}{\bar{\rho}}\left(\frac{\pd\ln \bar{p}}{\pd z}\right)^2\left(\frac{\pd
\bar{U}}{\pd z}\right)^{-2}\left(\nabla_{\rm ad}-\frac{d\ln \bar{T}}{d\ln \bar{p}}\right)\right],
\end{equation*}
where overbar denotes horizontal average.  The fastest and medium
solutions are convective near the top wall.  This is due to the higher
wind speed and energy injection, which cause a steeper temperature
gradient there.

\begin{deluxetable}{ccccccc}
\tablewidth{0pt}
\tablecaption{Viscosity, turbulent viscosity, relative importance of
  artificial bulk viscosity, relative importance of turbulent dissipation, Mach number,
  and Richardson number for a characteristic solution of each type in
Fig~\ref{eps.1g_vx}.}
\tablehead{\colhead{Solution type} & \colhead{${\rm log_{10}}\nu/c_sH_p$} & 
\colhead{$\nu_{\rm t}/c_sH_p$} & \colhead{$\langle\dot{E}_{\rm abv}\rangle_t/\langle\dot{E}_{\rm in}\rangle_t$} &
 \colhead{$\langle\dot{E}_{\rm turb}\rangle_t/\langle\dot{E}_{\rm in}\rangle_t$}
 &\colhead{$\mathcal{M}$} & \colhead{Ri}} 
\startdata
Fastest (triangle) & -2.17  & 0.006 & 0.07 & 0.43  & 7.6 & $<0$  \\ \hline
Medium (square) & -2.57  & 0.010 &  0.22 & 0.60  & 5.5  & $<0$  \\ \hline
Slow A (circle) & -3.03  & 0.015  & 0.13 & 0.79  & 2.3  & 0.033 \\ \hline
Slow B (star) & -2.40  & 0.017  &  0.12 & 0.67  & 2.9  & 0.024 \\ \hline
\enddata
\label{tab:eps.1g}
\end{deluxetable}

Fig.~\ref{eps.1g_vx} shows that the fastest triangle solution in the
laminar regime has wind speed scaling roughly as $1/\nu$, as we expect
from our tests in \S\ref{sec:tests}.  Further, the perfectly laminar
solution at $\nu/c_sH_p=-1.03$ has $Ri=0.36$, and we check that it
should be linearly stable as well.  From our runs at lower forcing we
determine that for the fastest (triangle) solution, the viscosity that
maximizes the wind speed scales in proportion to the forcing:
$\nu_{peak}/\cs H_p\propto\epsilon$.  The peak wind itself decreases
only marginally.  We can understand this behavior by assuming a
steady-state balance between laminar dissipation of the mean flow and
the forcing \eqref{eq:forcing}.  These solutions are able to reach
such a high wind speed because they are nearly laminar, until
turbulent dissipation kicks in at low $\nu$.

We map the solution space at lower forcing and thermal diffusivity and
verify that there are four different solutions in the regime
$Pr\lesssim 1$.  Specifically, we looked at $\chi/c_sH_p=0.094$ and
$\epsilon=.025, .05, \text{ and }0.1$, and also at $\chi/c_sH_p=
0.047,\epsilon=0.1$.  These parameters correspond to the four values
in the first three rows of table \ref{tab:mach}, discussed in greater
detail below.  There are small quantitative differences in the results
at $\chi/c_sH_p=0.047$, $\epsilon=0.1$.  The fastest solution in
particular has peak horizontally averaged Mach number $\sim 15\%$
higher.  From our linear analysis we expect little difference in the
results for any Peclet number in the range $Pe > 10$.  In our
simulations we are continually injecting energy onto our grid though,
and the diffusivity affects how this energy is redistributed.  We
attribute the differences in the nonlinear behavior for $Pe \sim
O(10)$ to the important dynamical effect of this internal energy
redistribution.

The main takeaway point from these runs at different forcing and
thermal diffusivity is that in the high-Reynolds-number limit there is
a unique solution, for which the direct viscous dissipation of the
mean flow is negligible compared to turbulent and shock dissipation.  Further, if
the grid resolution is insufficiently low, with $N_p \lesssim O(3)$,
we may lock onto an artificially fast solution, even at
$\nu/c_sH_p=0$.  We check the latter result for the range of
parameters we studied and not just for those relevant to Fig.~\ref{eps.1g_vx}.

We now attempt to extract an effective turbulent viscosity from the
physically relevant slow A solution.  First note that the energy
errors in Fig \ref{eps.1g_energy} are largest for the slow A
solutions.  In order to decrease the numerical dissipation we can
either increase the shear viscosity $\nu/c_sH_p$ or we can make the
grid resolution finer.  We cannot increase the shear viscosity
indefinitely because we will enter the multiple-solution regime
(Fig.~\ref{eps.1g_vx}) or the laminar viscosity will become large
enough to modify the properties of the solution.  We must balance this
against the increased computational cost of running the simulations at
higher resolution.  To calculate the turbulent viscosity we use
sufficiently high grid resolution and shear viscosity such that the
numerical energy dissipation rate is $\lesssim 5\%$ of the energy
injection rate, but shear viscosity is low enough such that the gross
properties of the solution are not modified.

In Tables \ref{tab:mach} and \ref{tab:turbvisc} we give the Mach
number and turbulent viscosity for a series of slow A solutions at
different horizontal forcing amplitudes and thermal diffusivities.
The exact value of the explicit viscosity we use is unimportant, so long as the
viscous dissipation length is larger than the grid scale and the
viscosity is small enough to leave the gross properties of the slow A
solution unmodified.  In general the runs satisfy $Re>10^3$.  The runs
at $\chi/c_sH_p=0.094$ and $\epsilon=0.025,\,0.01$ are done at the standard
resolution of $N_p=10$ grid points per pressure scale height.  The
rest are done at twice the standard resolution (i.e., twice as many
points in each dimension), with the exception of the run at
$\chi/c_sH_p=0.013$, $\epsilon=0.01$, which is done at three times the
standard resolution, and the run at $\chi/c_sH_p=0.094$,
$\epsilon=0.005$, which is done at 1.5 times the standard resolution.
We emphasize that these values of $N_p$ indicate the grid resolution
required at each set of parameter values to conserve energy within
$5\%$, not the resolution required to find the slow A solution at
$\nu/c_sH_p=0$.  At the standard resolution of $N_p=10$, we still find the slow A solution at
$\nu/c_sH_p=0$ for all parameters surveyed, but the fractional energy
error may be as large as $0.3$ for the lowest diffusivity runs.

\begin{deluxetable}{cccc}
\tablewidth{0pt}
\tablecaption{Mach number for a sequence of slow A solutions at different
  diffusivity and horizontal forcing.}
\tablehead{\colhead{$\epsilon$} & \colhead{} & \colhead{$\chi/c_sH_p$} & \colhead{} \\
\colhead{} & \colhead{.094} & \colhead{.047} & \colhead{.013} } 
\startdata
.1 & 2.3  & 2.3 & \\ \hline
.05 & 1.9  & & \\ \hline
.025 & 1.7  & & \\ \hline
.01 & 1.7  & 1.6  & 1.7\\ \hline
.005 & 1.5  & 1.7  & 1.7 \\ \hline
\enddata
\label{tab:mach}
\end{deluxetable}

\begin{deluxetable}{cccc}
\tablewidth{0pt}
\tablecaption{Turbulent viscosity $\nu_{\rm t}/c_sH_p$ for a sequence of slow A
  solutions at different diffusivity and horizontal forcing.}
\tablehead{\colhead{$\epsilon$} & \colhead{} & \colhead{$\chi/c_sH_p$} & \colhead{} \\
\colhead{} & \colhead{.094} & \colhead{.047} & \colhead{.013} }
\startdata
.1 & 0.015  & 0.017 & \\ \hline
.05 & 0.010  & & \\ \hline
.025 & 0.0061  & & \\ \hline
.01 & 0.0024  & 0.0023  & 0.0024\\ \hline
.005 & 0.0012  & 0.0011  & 0.0010 \\ \hline
\enddata
\label{tab:turbvisc}
\end{deluxetable}

The Mach number decreases gradually with forcing
amplitude $\epsilon$ for $\epsilon=0.1$ down to $\epsilon=0.01$, and is
roughly independent of $\epsilon$ for $\epsilon=0.005 - 0.01$.  It is
also independent of the diffusivity strength for $Pe \sim 10-100$, at
all forcing amplitudes surveyed.  The latter is consistent with
our linear analysis, and we attribute this result to the lower energy
injection rate as compared to the faster solutions from Fig.~\ref{eps.1g_vx}.
There is less thermal redestribution of heat in the slow A solutions.

\begin{figure}[htp]
\centering
\includegraphics[scale=.8]{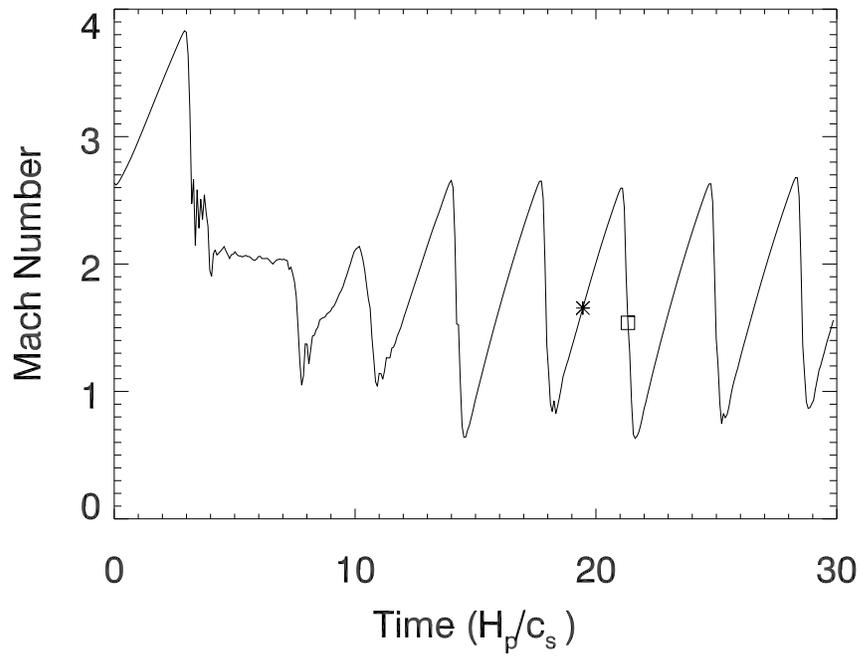}
\caption{Horizontally averaged Mach number as a function of time for
the slow A solution at $\chi/c_sH_p=0.013$, $\epsilon=0.01$.  The steady
state after time $\sim 10H_p/c_s$ alternates between phases of gradual
acceleration, and sharp drops of Mach number due to a violent
instability.  The star and box denote points at which we illustrate
the vector field in figs \ref{nu0001_chi001_velocity} and
\ref{nu0001_chi001_velocity_shock}.}
\label{nu0001_chi001_windspeed}
\end{figure}

\begin{figure}[htp]
\centering
\includegraphics[scale=.5]{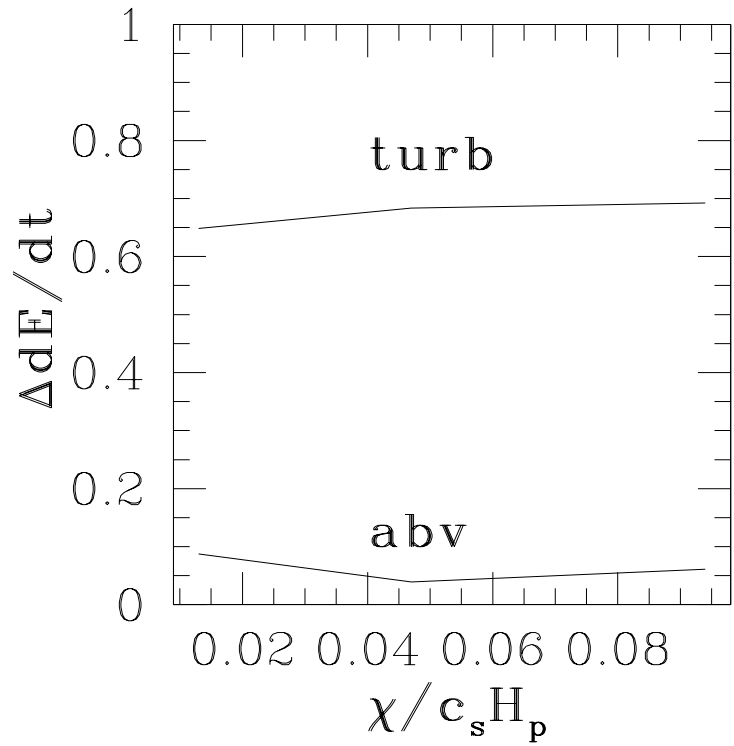}
\caption{Dissipation fractions via artificial bulk viscosity (abv)
and turbulence at $\epsilon=0.01$ and the
sequence of $\chi/c_sH_p$ from Table \ref{tab:mach}.  The turbulent
dissipation is largely irreversible compression, and it dominates over
the shock dissipation.}
\label{dissiprates}
\end{figure}

Fig.~\ref{nu0001_chi001_windspeed} shows the Mach number as a function
of time for the run at $\epsilon=0.01$ and $\chi/c_sH_p=0.013$.  The
run has reached a statistical steady state after time $\sim
10H_p/c_s$.  The steady state is characterized by a drive-up phase in
which the Mach number gradually rises, and a short slowdown phase in
which the Mach number rapidly drops.  In Fig.~\ref{dissiprates} we
show the relative contributions of dissipation due to artificial bulk
viscosity and turbulence, as defined in \S\ref{sec:diagnostics}.  Both
dissipation terms are confined to the slowdown phases.  Shocks appear
briefly during slowdown, but the majority of the kinetic energy
dissipation occurs in the absence of shocks.  We thus conclude that
shocks do not significantly contribute to the irreversible
compressional heating.

\begin{figure}[htp]
\centering
\includegraphics[scale=.7]{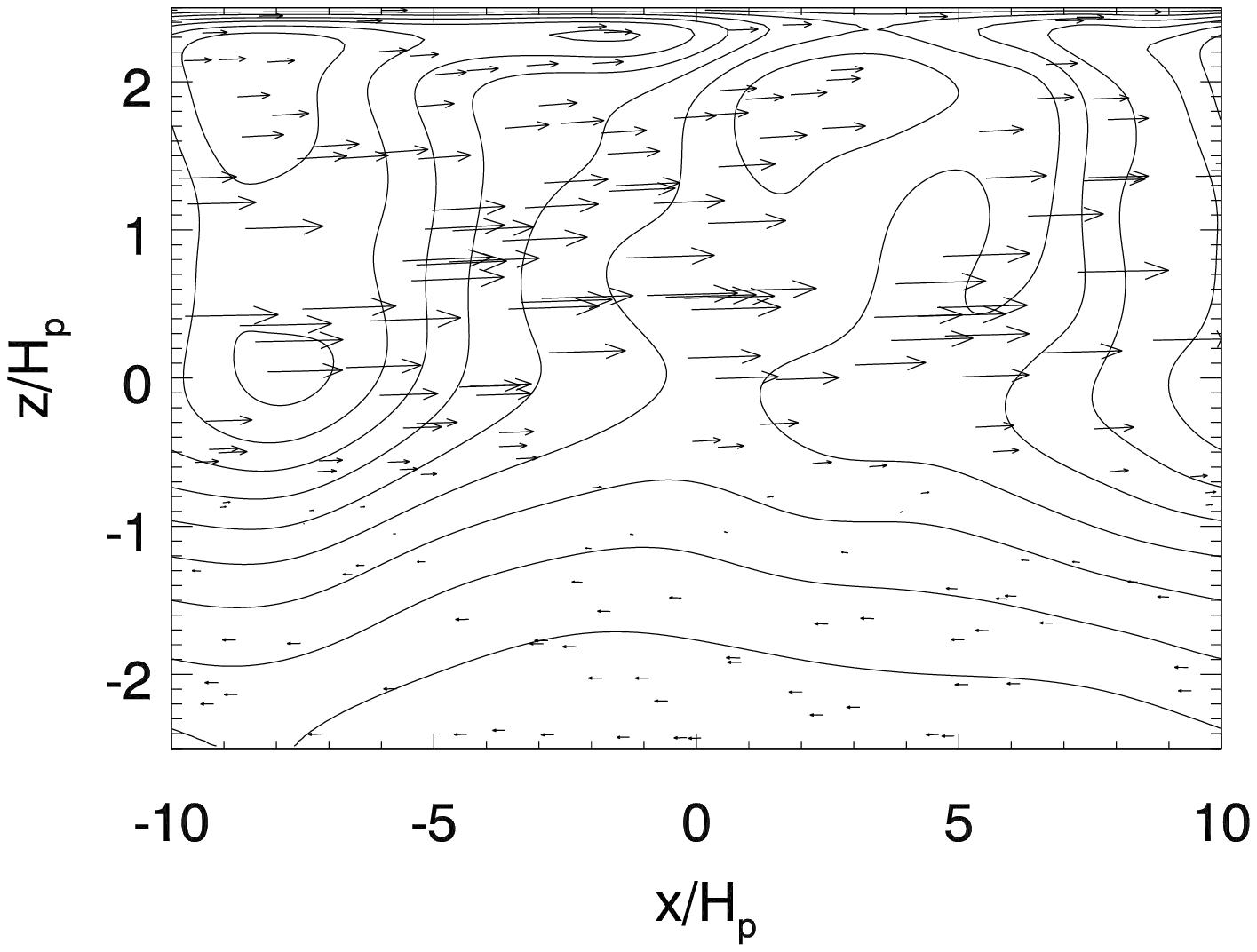}
\caption{Velocity field for the slow A solution at $\chi/c_sH_p=0.013$,
$\epsilon=0.01$ during the drive up phase (starred point in Fig.~\ref{nu0001_chi001_velocity}).
Equally spaced temperature contours are overlaid from $T_{\rm wall}$ to $T_{\rm
max}=1.07T_{\rm wall}$.  The minimum Richardson number is
$Ri=0.11$.}\label{nu0001_chi001_velocity}
\includegraphics[scale=.7]{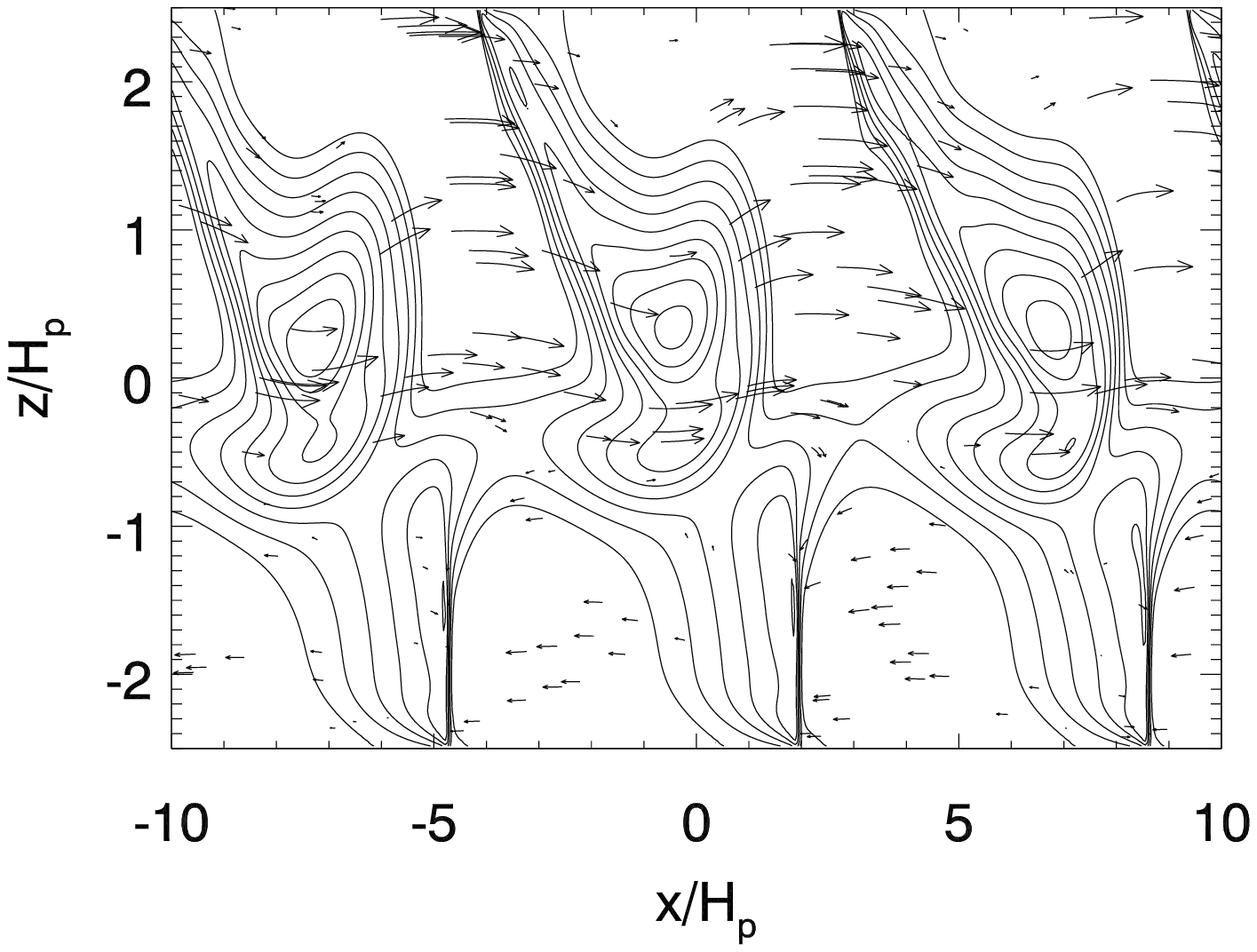}
\caption{Like Fig.~\ref{nu0001_chi001_velocity} but
during the unstable phase (boxed point in
Fig~\ref{nu0001_chi001_velocity}), and with
contours from $T_{\rm wall}$ to $T_{\rm max}=1.46T_{\rm wall}$.
Minimum $Ri=0.04$.  Periodic shocks occur at
$z/H_p<-1$, but the bulk of the heating is due to irreversible
compression.  }\label{nu0001_chi001_velocity_shock}
\end{figure}

We now characterize the nature of the sharp slowdown phases that limit
the wind speed in our flow.  The drive-up phase is
relatively smooth with small changes in the state variables across the
flow.  Fig.~\ref{nu0001_chi001_velocity} shows the velocity vector
field with temperature contours from $T_{\rm wall}$ to $T_{\rm
max}=1.07T_{\rm max}$ for the starred point from Fig.~\ref{nu0001_chi001_windspeed}.
At a short time before reaching the
peak wind speed, $t\approx 0.2 H_p/c_s$, a vigorous instability with 3
wavelengths in our box develops.
Fig.~\ref{nu0001_chi001_velocity_shock} shows the velocity vector field
with temperature contours from $T_{\rm wall}$ to $T_{\rm
max}=1.46T_{\rm max}$ for the boxed point from Fig.~\ref{nu0001_chi001_windspeed}.
The unstable mode is clearly visible
in the temperature profile roughly halfway through the slowdown phase.
Note the shocks at $z/H_p<-1$ and $x/H_p=-4.5,2,8.5$.
Following our previous
determination that the shocks are weak, we conclude that turbulent
mixing associated with this unstable mode is the primary agent for
dissipating kinetic energy and reducing the wind speed.

The dimensions of our box enforce a periodicity under the
transformation $x\to x+20H_p$.  If the unstable mode is an ordinary
Kelvin-Helmholtz mode, then from our linear analysis the only modes
that can grow have either $1,2, \text{ or } 3$ wavelengths in the box.
Apparently the mode with 3 wavelengths and wavenumber $k_xH_p\approx
.94$ is the fastest growing of the three.  The Richardson numbers
satisfy $Ri<1/4$.  We verify that the growth of this instability and its
effect on the wind speed are independent of the dimensions
of our domain.  We increase the horizontal extent of our box from
$L_x=20H_p$ to $L_x=30H_p$ and $L_x=40H_p$ and verify the presence of
this instability.  At $L_x=30H_p$ the fastest growing modes have 3 or
4 wavelengths in the box, and at $L_x=40H_p$ they
have 4 or 5 wavelengths.  These wave numbers are roughly
consistent with our linear analysis.  More importantly though, the
time-averaged Mach number of the flow remains unchanged.

\section{Discussion}\label{sec:discussion}

Ultimately we would like to provide a dissipation prescription for
global simulations that do not resolve the turbulent nature of the
wind profile.  We first note that from Fig
\ref{nu0001_chi001_windspeed} we find a horizontally and time averaged
Mach number $\mathcal{M}\sim 1.7$ and peak horizontally averaged Mach
number $\mathcal{M}\sim 2.7$.  The peak occurs just before the rapid
growth of instabilities, and so the latter corresponds to the absolute
peak Mach number in the flow.  Global simulations are often more
strongly supersonic.  For example, the model of HD~209458b by
\citet{Cooper_Showman05} finds $v_{\max}\approx 9\,{\rm km\,s^{-1}}$
at the 2.5~mbar level (see Paper~I for a survey of the calculations
prior to 2009).

Most of these studies of global circulation employ reduced equations
of motion based on the approximation of vertical hydrostatic
equilibrium: shallow-water, equivalent barotropic, or primitive
equations.  This is done in order to cope efficiently with the large
ratio of horizontal to vertical scales [eq.~\eqref{eq:scaleheight}],
but the approximations used also have the effect of filtering out
sound waves and shocks (even horizontally propagating ones).  Such
codes cannot directly represent the shocks and turbulent vertical
motions studied here, although they can represent horizontal shears
(which we do not) and hydraulic jumps
\citep[e.g][]{Showman_Fortney_etal09,Rauscher_Menou09}, which bear
some resemblance to shocks but do not conserve energy and may not be
triggered in the same circumstances.  Therefore no particular reason
exists as to why our results for the wind speeds should agree
quantitatively with those obtained from the primitive equations, for
example.

Comparison with the calculations carried out by
\citet{Burkert_etal05}, \citet{Dobbs-Dixon_Lin08} and \citet[herafter
DCL]{Dobbs-Dixon_Cumming_Lin10} is potentially more revealing, because
these calculations, like ours, use the full compressible hydrodynamic
equations.  The most recent of these studies included an explicit
shear viscosity.  It was found that the viscosity had little effect
unless at least $10^{10}{\rm\,cm^2\,s^{-1}}$, above which the
speed decreased monotonically.  For reference, $\nu=
10^{10}{\rm\,cm^2\,s^{-1}}$ corresponds to $Re\approx 600$ referred to
the sound speed and scale height.  
We find, however, that the wind
speed is not monotonic or even single-valued in the viscosity
(Fig.~\eqref{eps.1g_vx}).

We would like to understand what accounts for the difference between our
results and those of DCL.  
Compared to our body force and constant diffusivity, the thermal forcing and
radiative transfer implemented by DCL is much more sophisticated and
realistic, but it is hard to see what this has to do with the viscous
effects, especially since the behavior seen in Fig.~\eqref{eps.1g_vx}
exists for all choices of forcing and diffusivity that we have tried.
Furthermore, our direct forcing is usually stronger than DCL's
thermal forcing since, for example, a horizontal acceleration of
$f=0.025\,g$ would result in a Mach number $\sqrt{2\pi R_p
  f}/\cs\gtrsim 7$ after going half way around the planet without
dissipation.  And yet our Mach numbers are at most
comparable to those of DCL, the latter peaking at $\sim 3$ for low
viscosity.

More likely the difference in results has to do either with the
difference in the dimensionality and geometry of the calculations (3D
and global for DCL; 2D and local for us), or with spatial resolution.
It is well known that subsonic turbulence behaves differently in two
and three dimensions; however, since two dimensions usually favor
energy on large scales, our wind speeds might have been expected to
exceed DCL's if dimensionality were the most important factor.
Concerning numerical resolution, it appears that DCL's standard value is
$\lesssim 4$ cells per scale height, compared to our 10.
Since we and DCL use similar numerical methods, their calculations are
probably more diffusive than ours.   When we use a resolution
comparable to theirs, our wind speeds actually increase for the same
forcing (\S\ref{subsec:results}).  The fact that we and they find
comparable Mach numbers at our highest respective Reynolds numbers
suggests that our stronger forcing is partly offset by our finer resolution. 

A limitation of our local, two-dimensional models is the somewhat
arbitrary assumption that the nonpotential horizontal force is
compensated over a depth range $\Delta z\approx 2H_p$.  Other things
being equal, one would expect the flow speed to scale as $(\Delta
z)^{1/2}$ at constant forcing amplitude if Richardson number
rather than Mach number is the controlling parameter.  Inspection of
DCL's figures suggests that the horizontal speed scales approximately
linearly in $\ln p$ over the range $10^{-4}\mbox{-}10^0{\rm bar}$.
The sign change is not always obvious in their velocity plots, although
they infer the existence of return flow from their temperature profiles.
However, since the momentum density $\rho\bar v_x\propto
v_x(z)\exp(-z/H_p)$, this is much better localized around its sign
change than is $v_x$ itself.

We can use our turbulent viscosities to estimate a turbulent
diffusivity for composition, $K_{\rm t}$, as may be required to keep
the upper atmosphere turbulently mixed
\citep{Spiegel_Silverio_Burrows09}.  The Schmidt number, defined by
$Sc_{\rm t}=\nu_{\rm t}/K_{\rm t}$, in general has a complicated
dependence on flow properties \citep[see e.g.,][]{Huq_Stewart2008}.
For $Ri \lesssim O(10^{-1})$, however, $Sc_t$ is independent of $Ri$
and of order unity.  The data in Table \ref{tab:turbvisc} then gives
us mass diffusivities pertaining to the shear layers in Hot Jupiter
atmospheres: $K_{\rm t}\gtrsim10^{-3}\cs H_p\sim
10^{10}{\rm\,cm^2\,s^{-1}}$.  \citet{Spiegel_Silverio_Burrows09}
estimated that $K_{\rm t}\sim 10^7-10^{11}{\rm\,cm^2\,s^{-1}}$ is
required,  depending on grain size,  to maintain TiO as an optical
absorber at $p\lesssim 1 {\rm\,mbar}$.
Our values are toward the upper end of this range, but only in the
shear layer ($\sim 2H_p$).  To estimate $K_{\rm t}$ elsewhere, we
would need to estimate how much of the mechanical power dissipated in
the shear layer can propagate upwards and downwards in the form of
internal waves or shocks.  This we have not attempted.

\section{Conclusions}\label{sec:conclusions}

We have investigated the balance between horizontal forcing and mechanical
dissipation in model shear flows designed to imitate thermally driven
winds of jovian exoplanets.
Our main conclusions are as follows.
\begin{enumerate}
\item Following Paper I, some form of mechanical dissipation is
  required to offset the production of mechanical energy by the
  longitudinal entropy gradient.   Plausible dissipation mechanisms
  include shocks, turbulence, and MHD drag.

\item Entropy stratification has a stabilizing influence on Kelvin-Helmholtz (KH)
  modes, but  thermal diffusivity ($\chi$) tends to undercut this
  influence to a degree that can be quantified
  by the Peclet number $Pe\equiv\cs H_p/\chi$ based on the sound speed
  ($\cs$) and pressure scale height ($H_p$).  We estimate $Pe\sim
  O(10^3)$ at the altitude where the thermal timescale is comparable
  to the circumferential flow time at Mach~1, but  $Pe\lesssim O(1)$
  near the infrared photosphere.

\item Linear analysis indicates that for $Pe\gtrsim 10$, thermal
  diffusion has little effect on the stability of KH modes whose
  horizontal wavelength is comparable to the vertical thickness of the
  shear layer; the usual adiabatic Richardson criterion $Ri\equiv N^2/(\partial
  U/\partial z)^2>1/4$ is sufficient for the stability of these
  short-wavelength modes.  For typical conditions on the day sides of
  hot Jupiters, $Ri<1/4$ implies transsonic or supersonic winds,
  presuming that the shear layer is no narrower than a pressure scale height.

\item Nevertheless, at any finite $Pe$ and $Ri$, inviscid flows with an
  inflection point in the shear profile are unstable at sufficiently
  long horizontal wavelengths, at least when the flow is confined
  within a channel of finite vertical extent.  We doubt that these
  long-wavelength KH modes are important for hot-Jupiter winds because
  their growth rates are very small (inversely proportional to
  wavelength) and because the planetary circumference is not
  large enough compared to $H_p$ to accommodate them at
  $Pe\gtrsim 10^3$.
 
\item We estimate a turbulent diffusivity for composition in hot
  Jupiter atmospheres.  For $Pe\approx 100$, we find $K_{\rm
  t}/c_sH_p\approx 10^{-3}$, depending on the strength of the
  horizontal forcing.  However, these values pertain to regions of
  maximum vertical shear; turbulent diffusivities may be much smaller
  elsewhere.

\item Nonlinear, two-dimensional (azimuth and depth), horizontally
  forced local simulations saturate at time-averaged Mach numbers
  $\sim 2$  when the Reynolds number $Re\equiv \cs H_p/\nu\gtrsim
  10^3$.  Given the widths of our shear layers ($\sim 2H_p$) and background
  stratification, this corresponds to minimum Richardson numbers
  slightly below the adiabatic critical value of $1/4$.
  In this regime, which is relevant to hot Jupiters, we find that the
  saturated wind speed depends only weakly on $Pe$ and on the strength
  of the forcing.

\item Dissipation in the high-$Re$ regime is due mainly to the work
expended to overturn the stratification rather than shocks, though
weak shocks are present.  It is highly intermittent, at least in our
two-dimensional calculations, and appears to be triggered by
a recurrent linear instability of Kelvin-Helmholtz type.

\item At $Re\lesssim 10^3$ or at low numerical resolution
  (significantly fewer than 10 cells per pressure scale height),
  multiple sequences of solutions appear in overlapping ranges of
  Reynolds number, differing in their wind speed and predominant
  dissipation mechanisms.  The fastest have Mach numbers at least
  twice as large as what we believe to be the correct inviscid values
  for our forcing strengths.

\end{enumerate}

The last three points suggest that some global simulations of hot
Jupiters may have overestimated wind speeds due to incomplete physics
(e.g. neglect of vertical accelerations and sound waves) and/or
insufficient spatial resolution.  Pending sufficient computational
resources to resolve the physical dissipation mechanisms directly in
global simulations, we suggest that artificial dissipative terms be
added to the equations of motion so as to prevent strongly supersonic
shears and small Richardson numbers.  The appropriate form for these
terms will depend upon the equation sets and algorithms used, but
presumably it is desirable that they conserve energy and momentum, and
probably that they have a sharp threshold in Mach number or Richardson
number.

\acknowledgments

We thank Jim Stone for help with ZEUS.
This work was supported in part by the National Science foundation
under grant AST-0707373.

\goodbreak
\bibliographystyle{apj}
\bibliography{circ}


\newpage
\appendix
\section{Exact linear solutions}

We provide here four limits of eqns \ref{eq:NavierStokeslin} and
\ref{eq:thetaeqnlin} for which there analytic solutions.  We test our
code against these solutions.

Taking $N^2=0,\nu \to 0$, eqns \ref{eq:NavierStokeslin} and
\ref{eq:thetaeqnlin} admit solutions satisfying the inviscid
perturbation equation with vanishing temperature perturbation,
\begin{align}
v_{1z}''&=\left(-\frac{k_xU''}{\sigma}+k^2\right)v_{1z},\\
\theta_1&=0,
\end{align}
in which $k\equiv(k_x^2+k_y^2)^{1/2}$.
For a hyperbolic tangent velocity profile $U=\tanh(z)$, the inviscid
perturbation equation for the marginal $\omega=0$ mode becomes
\begin{equation}
v_{1z}''=-\left(2\sech^2z\,-k^2\right)v_{1z}.
\end{equation}
A solution to this equation that vanishes as $z\to\pm\infty$ is
\begin{equation}
v_{1z}=\sech z\,,\qquad k^2=1.
\end{equation}

In the limit
\begin{equation}
\label{limits1}
U=0,\nu \to 0,N^2=0,
\end{equation}
eqns \ref{eq:NavierStokeslin} and \ref{eq:thetaeqnlin} become
\begin{equation}
\label{eq:NavierStokeslinlimit}
v_{1z}''=(k^2)v_{1z}-\frac{i}{\omega}{k^2}\theta_1,
\end{equation}
\begin{equation}
\label{eq:thetaeqnlinlimit}
\theta_1''=\left(-\frac{i\omega}{\chi}+k^2\right)\theta_1,
\end{equation}
which for $Re(\omega)=0$ and $Im(\omega)< -(k^2)\chi$ admit
the decaying mode solutions
\begin{equation}
\theta_1=A\cos\left[\left(-k^2+\frac{i\omega}{\chi}\right)^{1/2}z +\phi \right],
\end{equation}
\begin{equation}
v_{1z}=A\frac{\chi k^2}{\omega^2}\cos\left[\left(-k^2+\frac{i\omega}{\chi}\right)^{1/2}z +\phi \right]+Be^{{k}z}+Ce^{-{k}z}.
\end{equation}

In the limit $\chi \to \infty$ with conditions \ref{limits1}, we
obtain solutions to eqns \ref{eq:NavierStokeslinlimit} and
\ref{eq:thetaeqnlinlimit} of the form
\begin{equation}
\theta_1=Ae^{{k}z}+Be^{-{k}z},
\end{equation}
\begin{equation}
\begin{split}
v_{1z}=\frac{i}{4\omega}\left(Ae^{{k}z}+Be^{-{k}z}+2{k}z\left(Ae^{{k}z}-Be^{-{k}z}\right)\right)\\
+Ce^{{k}z}+De^{-{k}z}.
\end{split}
\end{equation}

Finally, in the limit
\begin{equation}
U=0,\chi \to 0, \nu \to 0,
\end{equation}
eqns \ref{eq:NavierStokeslin} and \ref{eq:thetaeqnlin} combine to give
\begin{equation}
\label{eq:NavierStokeslinlimit2}
v_{1z}''={k^2}\left(1-\frac{N^2}{\omega^2}\right)v_{1z}.
\end{equation}
For $\omega$ real and $N^2/\omega^2 > 1$, eqn
\ref{eq:NavierStokeslinlimit2} has solutions of the form
\begin{equation}
v_{1z}=A\cos\left[k\left(\frac{N^2}{\omega^2}-1\right)^{1/2}z+\phi\right],
\end{equation}
\begin{equation}
\theta_1=A\frac{iN^2}{\omega}\cos\left[k\left(\frac{N^2}{\omega^2}-1\right)^{1/2}z+\phi\right].
\end{equation}

\section{Long-wavelength modes}

Here we demonstrate analytically for the Boussinesq equations that there exist long-wavelength instabilities
at arbitrarily large Richardson number if the Peclet number is
finite.  It is helpful to adopt dimensionless variables.  Supposing
$U(z)\approx \pm U_0$ at $|z|\gg H$, where $H$ is the width of the
shear later, we define these variables as follows:
\begin{align}
  \label{eq:dimlessvars}
  z_* &= z/H & U_*(z_*) &= U(z)/U_0  \nonumber\\
v_*(z_*) &= v_{1z}(z)/U_0 & \theta_* &=  \frac{\theta_1}{N^2H}  \nonumber\\
 k_* &= H\sqrt{k_x^2+k_y^2} & \cos\alpha &= k_x/\sqrt{k_x^2+k_y^2} & c_* &= \omega/(k_xU_0)\nonumber\\
Re &= U_0H/\nu & Pe &= U_0H/\chi & Ri &= (NH/U_0)^2\,.
\end{align}
Equations \eqref{eq:NavierStokeslin} \& \eqref{eq:thetaeqnlin} become,
with primes denoting $d/dz_*$,
\begin{subequations}\label{eq:dimlesseqns}
  \begin{align}
    \label{eq:dimless1}
   \left[v''_*+\frac{U''_*}{c_*-U_*}v_*-k_*^2v_*\right] (c_*-U_*)\cos\alpha
-\frac{i}{k_*Re}\left(\frac{d^2}{dz_*^2}-k_*^2\right)^2v_*
&= -Ri\, k_*\theta_*,\\
\theta''_*+\left[i Pe\,k_*\cos\alpha\,(c_*-U_*)-k_*^2\right]\theta_* &= Pe\, v_*.
  \end{align}
\end{subequations}

We take the inviscid, long wavelength limit in such a way that $k_*Re\to\infty$ while $k_*\to 0^+$, with $Pe$ and $Ri$
regarded as finite and nonzero.  The wavelength $2\pi|\boldsymbol{k}|^{-1}$ is not infinite, but rather $\gg 2\pi H$, so that
we may neglect the explicit appearances of $k_*$
in the equations above yet regard $c_*$ as finite.
If $c_*$ should turn out to have an imaginary part, then the growth
rate $\mbox{Im}(\omega)\to k_xU_0\mbox{Im}(c_*)$ as $k_*\to0$.
The potential temperature $\theta_*$ decouples:
\begin{subequations}\label{eq:dimzeroth}
  \begin{align}
    \label{eq:vstareqn}
   v''_*+\frac{U''_*}{c_*-U_*}v_* &=0\,,\\
    \label{eq:thetastareqn}
\theta''_* &= Pe\, v_*.
  \end{align}
\end{subequations}
The first of equations \eqref{eq:dimzeroth} is the long-wavelength limit of the inviscid, unstratified Kelvin-Helmholtz
problem with a continuous shear profile $U_*(z_*)$.  Given boundary conditions $v_*(\pm z_{*,\max})=0$
for some $z_{*,\max}$ that is finite and sufficiently large,
and presuming that $U_*$ is an odd function of $z_*$ so that $U''_*(0)=0$, 
then there is a solution to $\eqref{eq:dimzeroth}$ for which $c_*$ has a positive imaginary part.  In fact,
although it doesn't satisfy the boundary conditions, $v_*\propto U_*-c_*$ is a solution to eq.~\eqref{eq:vstareqn}
for any $c_*$, and a second solution $v_*=f(z_*)(U_*-c_*)$ can be found by solving a first-order equation for $f'(z_*)$,
followed by a quadrature.  Requiring that this solution vanish at both $\pm z_{*,\max}$ determines $c_*$ via
\begin{equation}
  \label{eq:ceqn}
  \int_{-z_{*,\max}}^{z_{*,\max}}\frac{dz_*}{[U_*(z_*)-c_*]^2}=0.
\end{equation}
Since $U_*$ is odd and $\approx\pm 1$ at large $|z_*|$, it follows easily that $c_*\approx\pm i$ when
$z_{*,\max}\gg 1$.  

This demonstrates the existence of inviscid instability when
$|\boldsymbol{k}|^{-1}\gg z_{*,\max}^2\gg 1$.
If $z_{*,\max}\gtrsim k_*^{-1/2}$,  however,  then a separate analysis would be needed, because the solution of
eq.~\eqref{eq:thetastareqn} might yield $\theta_*\sim Pe\,z_{*,\max}^2v_*$, and then the righthand side
of eq.~\eqref{eq:dimless1} could be $O(1)$ or larger.  We have not done this analysis because the numerical
evidence indicates that the long-wavelength growth rates are too slow to be important for our
applications.

\end{document}